# Venus Long-Life Surface Package (VL2SP)[1]

A response to ESA's Call for New Scientific Ideas, September 2016

Colin Wilson[1], Carl-Mikael Zetterling[2], W. Thomas Pike[3]

Atmospheric, Oceanic, and Planetary Physics, Oxford University, UK (colin.wilson@physics.ox.ac.uk);
[2] KTH Royal Institute of Technology, Sweden, [3] Electrical and Electronic Engineering, Imperial College London, UK.

## Executive Summary

Measurements in the atmosphere and at the surface of Venus are required to understand fundamental processes of how terrestrial planets evolve and how they work today. While the European Venus community is unified in its support of the EnVision orbiter proposal for the M5 opportunity, many scientific questions also require in situ Venus exploration. ESA has already explored Venus entry / descent probe science in its Planetary Entry Probe (PEP) study [ESA PEP study, 2010], and Venus balloon science in its Venus Entry Probe Study [ESA VEP study, 2005]; Venus balloons were also explored in detail by the European Venus Explorer (EVE) M1/M2 and M3 proposals [Chassefiere et al., 2009; Wilson et al., 2012]. While those in situ mission concepts remain scientifically compelling and technically feasible, the present call requests **new** scientific concepts. Therefore, in the present document, we suggest **a long-duration lander at Venus, which would be capable of undertaking a seismometry mission, operating in the 460°C surface conditions of Venus.**

Venus is arguably the Earth's closest twin. It is almost the same size as Earth, and probably formed at the same time as Earth with apparently similar bulk composition. However, its evolution has differed from that of Earth; it apparently lost most of its primordial water leaving a massive carbon dioxide atmosphere and enormous greenhouse effect, giving it its high surface temperature of 460°C. However, the interior structure of the planet are completely unknown, other than its mean density. Radar maps have shown Venus to be covered with volcanic and tectonic features, and mounting evidence, including observations from Venus Express, suggests that some of these volcanoes are active today. Assessing Venus' current seismicity, and measuring its interior structure, is essential if we are to establish the geological history of our twin planet, for example to establish whether it ever had a habitable phase with liquid water oceans. Although some constraints on seismic activity can be obtained from orbit, using radar or ionospheric observation, as will be detailed below, the most productive way to study planetary interiors is through seismometry.

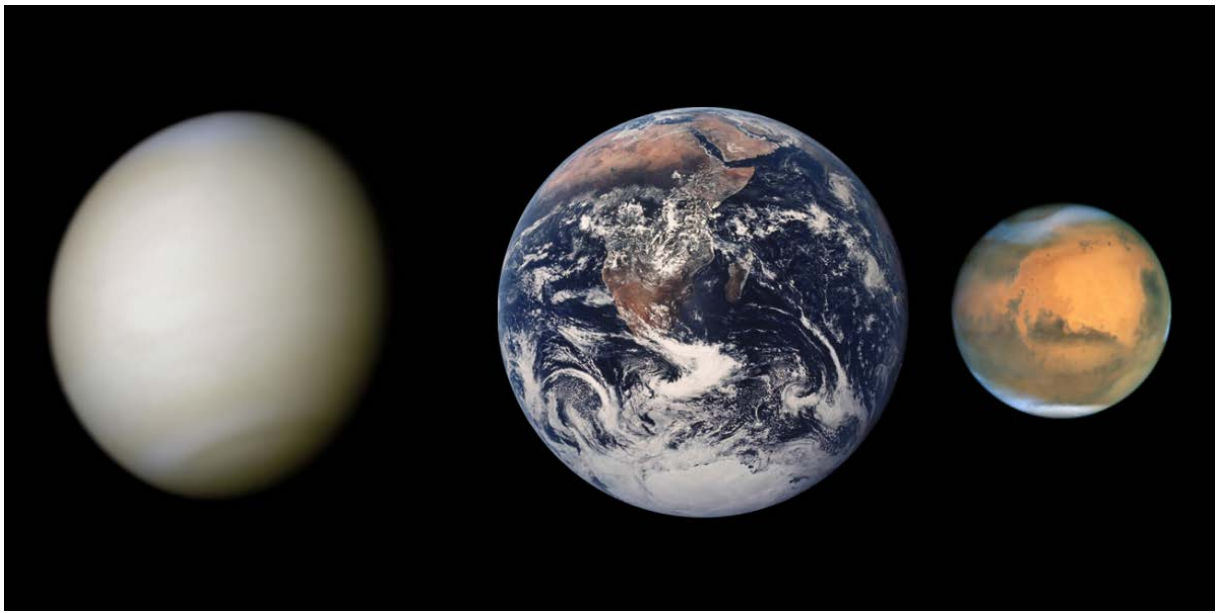

---

[1] Note: The title of the proposal given in the Letter of Intent was "Venus In-Situ Mission (VISM). The present proposal has changed in name to reflect changed priorities and to avoid confusion with an earlier "Venus Interior Structure Mission (VISM)", led in 1993 by Ellen Stofan et al.

Seismometry requires a mission duration of months or (preferably) years. Previous landers have used passive cooling, relying on thermal insulation and the lander's thermal inertia to provide a brief window of time in which to conduct science operations – but this allows mission durations of hours, not months. Proposals relying on silicon electronics require an electronics enclosure cooled to < 200 °C; the insulation, cooling and power system requirements escalate rapidly to require a > 1 ton, > €1bn class mission, such as those studied in the context of NASA flagship missions. However, there are alternatives to silicon electronics: in particular, there have been promising advances in silicon carbide (SiC) electronics capable of operating at temperatures of 500°C. For the post-2030 timeframe addressed in this call, it will be possible to assemble at least simple circuits using SiC components, sufficient to run a seismometry lander. SiC components for sensors, CPUs, memory, and telecommunications systems are all being developed. Power sources remain difficult at the surface of Venus; only a few W / m2 of sunlight reach the surface of Venus, due to the thick cloud layer, rendering solar power impractical; therefore we rely on a radioisotope thermoelectric generator to provide ~ 25 W of electrical power needed to power the lander's electronics.

We are proposing a Venus Long-Lived Surface Package (VL2SP) consisting of power source (RTG), science payload (seismometer and meteorology sensors), and ambient temperature electronics including a telecommunications system. We do not specify how this VL2SP gets to the surface of Venus: it might be carried to the surface of Venus by a larger Venus probe/lander, or it may enter with a dedicated entry shell and descent/landing system. We estimate that an orbiter providing data relay would be essential for a small (<100 kg class) VL2SP; a full mission study of the VL2SP concept should either include a data relay orbiter or assume that a Venus orbiter with this capability is available in a post-2030 timeframe for which this concept is being considered.

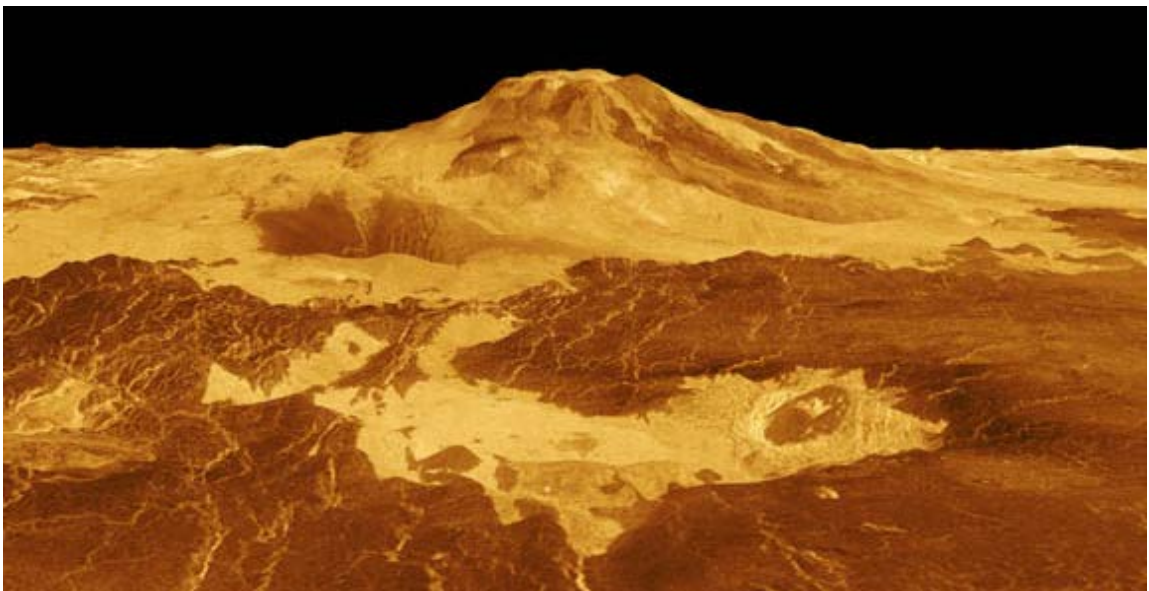

**Figure 1 - Venus's Maat Mons volcano, revealed by NASA's Magellan radar.
Is Venus alive or dead, volcanically and tectonically? Why does it not have a geodynamo?**

This proposal borrows extensively from a recent study conducted at the Keck Institute of Space Studies [KISS report, 2015]. It is also informed by a large research programme entitled "Working on Venus" being undertaken at KTH Royal Institute of Technology, Sweden under the leadership of Carl-Mikael Zetterling, which has the aim of developing SiC components for a Venus lander

# 1 Introduction & Science rationale

The formation, evolution, and structure of Venus remain a mystery more than 50 years after the first visit by a robotic spacecraft. Radar images have revealed a surface that is much younger than those of the Moon, Mercury, and Mars, covered with volcanic and tectonic features, many of which are quite unlike those generated by plate tectonics on Earth. The style of volcanism and tectonism on Venus today are not known. Is it still losing appreciable amounts of heat from its interior, as is Earth? Is this heat lost through the crust, by conduction, or is it lost through volcanism and/or tectonism?

The history of geological evolution, too, is not known. Does volcanism occur continuously at a low rate, in what is known as the stagnant lid model? Or are there catastrophic and repeated episodes of widespread volcanic eruption, in which large fractions of the planet are resurfaced? No signs of crustal subduction zones were found in the Magellan dataset, but there are many other signs of tectonic activity, such as extension fractures and wrinkle ridges; what role does tectonism play in allowing heat to be released from the interior? As on Earth, Venus' volcanoes come in many different shapes and sizes. There are steep-sided domes, suggesting very viscous lavas; channels which run for hundreds of kilometres, suggesting very low viscosity lava; and also some features which may be associated with pyroclastic flows.

In addition to constraining current mechanisms of heat loss on Venus, seismology would help to constrain the interior structure of Venus, and help constrain its mantle evolution style. Analysis of Venus' topography and gravity signatures has revealed nine highland regions thought to lie atop upwelling mantle plumes. Although this apparently shows that the mantle is convecting, Venus does not have a geodynamo, generating an intrinsic magnetic field. Is Venus' mantle convection organised in such a way that no geodynamo results? Or is its crust electrically conductive enough that any magnetic field generated by the internal convection remains within the crust? The magnetic history of Venus will have had a great influence on the history of its interaction with the solar wind; in particular, an early loss of the geodynamo may have increased water loss from Venus' early atmosphere. Clearly, the history of Venus' mantle convection and its magnetic field have important consequences for its evolution and habitability.

The high atmospheric density at the surface of Venus – around 50 kg m$^{-3}$ [Seiff et al., 1985] – has the result that surface-atmosphere coupling is much more efficient on Venus than on Earth. This presents both opportunities and challenges for seismometry. To start with the positive aspects: If the ground shakes due to seismic activity, pressure waves can propagates upwards from the surface as infrasonic waves (frequencies typically in the 0.01 to 10 Hz range), continuing upwards into the lower thermosphere where their amplitude increases and the waves break. This offers opportunities for detection of seismic activity either using infrasound detectors on balloon platforms, or by detecting temperature fluctuations in the ionosphere from a dedicated camera on an orbiter. The second of these techniques – seismic mapping from orbit – could be carried out simultaneously with a surface mission and so will be discussed again in Section 5. However, these "remote sensing" techniques for observing seismic activity may be difficult to interpret, because the seismic signal propagates through the atmosphere before being detected; therefore, for the present proposal we will focus on the surface seismometer element.

The high atmospheric density also presents a problem for seismometry, in that oscillations in the atmosphere such as wind gusts are readily transmitted into the soil. Therefore, even a buried seismometer may be sensitive to fluctuations in wind speed, through the effect of fluctuating wind stress exerted on the surface of the planet. Potentially, this sets a relatively high "noise floor" for the seismometer. Winds near the surface of Venus have been measured only rarely, but these have all shown wind speeds to be typically 0 - 1 m/s in the lowest 10 km of the atmosphere [Kerzhanovich & Limaye, 1985]. Although these winds are light, the high atmospheric density means that exert a high surface wind stress on the surface. The seismometer emplacement must therefore be designed to minimise wind noise. One approach to this is to have the seismometer directly emplaced on the ground with a wind shield over it, an approach which will be used at Mars by the InSight seismometry mission; however, this implies quite a complex lander, to allow emplacement of a seismometer + shield. A simpler approach may simply be to have a structure of low height, in order to not present a large cross section to the wind. In either case, care should also be taken to ensure that the structure is well coupled to the ground and will not rock or sway in response to wind fluctuations.

How many seismic events will VL2SP record? This will depend on the seismicity of Venus; on the sensitivity of the accelerometer; and on the noise level introduced by wind gustiness. Lorenz (2012) calculated that a seismometer with a 10 nm position measurement accuracy might be expected to detect roughly one event per Earth day, assuming levels of seismicity similar to those found on Earth; therefore a mission of at least 100 days (1/2 Venus year) would be required to build up a catalogue of 100 events. However, it is stressed that these are very much "rough order of magnitude" estimates based on seismicity data from Earth rather than Venus.

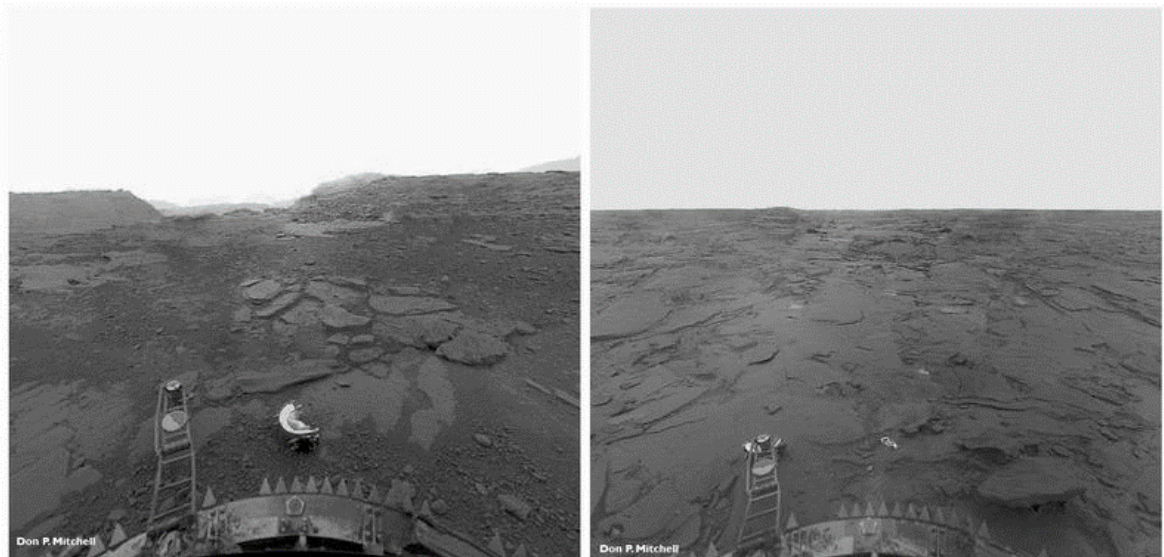

**Figure 2 - Venus's surface as seen from the Venera 13 lander on March 1 1982, reconstructed by Don Mitchell. How thick are these lava flows, and the crust beneath them? How old are they?**

## 2 Scientific payload

We present here a core payload which is essential for achieving the seismometric science goals of the mission; and additional sensors which would provide complementary science.

### 2.1 Seismometer

Fundamentally, a seismometer consists of a proof mass, whose position or velocity are measured. Velocity sensing is the principle used in geophones, the simplest version of which is just a magnetic proof mass moving within a coil; but velocities decrease with frequency, so position sensing is used for higher sensitivity at low frequencies. Capacitive inertial sensors are especially attractive for high-temperature applications because of the inherently low temperature dependence of the capacitive sensing principle.

Broad-band seismometry using a silicon sensor has been demonstrated with the flight-delivered sensors for the 2018 InSight mission to Mars [Pike et al. 2014]. The three micromachined sensors are shown in fig. 3 prior to integration. These sensors are micromachinined from single-crystal silicon by through-wafer deep reactive-ion etching to produce a suspension and proof mass with a fundamental vibrational mode of 6 Hz. The motion of the proof mass is sensed capacitively between an array of parallel electrodes on the proof mass and a matching fixed array on a glass strip separated by a fixed gap from and soldered on to the proof mass. The seismometer is robust to high shock (> 1000 g) and vibration (> 30 g rms). In addition, all three axes of the microseismometer deliver full performance over a tilt range of ±15 degrees on Mars, allowing operation after deployment without levelling. The fabrication of the wafers for these sensors is already compatible with a 500C environment. It is the packaging of the sensors that would have to be adapted, and

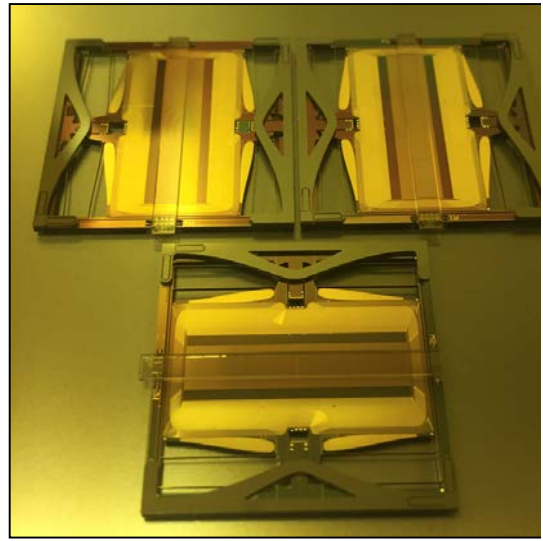

**25 mm**

**Figure 3 -** The microseismometer sensor set delivered for the InSight Mars lander

two routes are possible, replacement of the solder interconnects with thermal-compression binding, or use of high temperature solder.

The current sensors use magnetic feedback to provide a flat response in velocity. However, they have also been successfully run in an open-loop configuration and as this approach requires no magnets, it would be more suitable for Venus. The feedback electronics can also be much simplified for open-loop operation. A square-wave drive to the capacitance transducer on the sensor can be sensed using a single differential amplifier, demodulated with two switches and simple logic from the drive. Alternatively, the output from the preamplifier can be directly mixed onto a carrier signal for direct AM transmission. In both cases, only a small number of simple components are required.

The ability of the sensor to work under a range of tilts allows for direct deployment on the surface of Venus, either as a low profile surface package, or a hard or soft penetrator. For the latter, techniques have been developed that allow protection of the suspension up to 15,000 g [Hopf et al 2010], although these high shock loads are unlikely to be achieved in a Venus lander.

Operation of a capacitive accelerometer at high temperatures of up to 460°C and feasibility of capacitive detection at these temperatures was recently demonstrated at KTH Sweden [Asiatici et al,. 2016]. For these experiments, a bare silicon MEMS sensor die from a commercial single-axis capacitive accelerometer module was supplied by Colibrys (part of the Safran Group). While the accelerometer proved operational up to at least 460°C (the limit of our experimental set-up), decomposition of the polymer-based antistiction layer covering the proof-mass was observed, resulting in stiction problems, starting at a temperature of 400°C. This reliability issue is not considered a fundamental problem as the polymer-based antistiction coating can in principle be replaced by high-temperature stable antistiction coatings, e.g. nano-crystalline graphite coatings.

KTH teams (Prof. Zetterling, Prof. Östling and Prof. Niklaus) are currently developing SiC-based high-temperature electronics for open-loop signal read-out of the capacitive accelerometer, and to demonstrate a complete sensor unit that can operate at temperatures of up to 450-500°C. In a second step, a closed-loop signal read-out circuit is anticipated, which will most likely be required to achieve improved module sensitivity.

Colybris and other companies have announced that they are developing improved seismic-grade capacitive MEMS accelerometers. Thus, capacitive MEMS accelerometer technology together with SiC-based read-out electronics seems a very promising approach for ground-based and long-term seismic measurements in the context of a Venus mission. Ongoing developments of commercial capacitive accelerometers for seismology, mainly targeted at the oil prospecting industry, will be leveraged to enable further progress in this field.

Power requirements would be modest; a seismometer based on the Insight SEIS-SP would require around 200 mW in total, with a total package size for a three-axis seismometer of 30 x 30 x 30 mm. Data requirements can be adapted depending on the mission resources; sampling rate should ideally be 10 Hz but can be reduced as far as 1 Hz for a threshold mission. This allows us to define a baseline data rate requirement based on 1 sample per second from 3 seismometer axes, each sampled at 16 bits per second, i.e. 48 bits per second. Sampling at 10 Hz would push this up to hundreds of bits per second. This is the sizing requirement for the communications system.

## 2.2 Meteorological sensors

As was discussed above in Section 1, fluctuating winds can cause strain in the planetary surface, resulting in noise in seismic signals. Pressure variations, if present, would also cause spurious seismic signals. Wind and pressure sensors would help identify these sources of noise in the seismometer dataset. Air temperature measurement would also assist the interpretation of the seismic data; the temperature difference between day and night is thought to be less than 1 degree, but measurement of this day-night temperature difference would have significant implications not only for surface thermal stresses but also for atmospheric circulation.

There is significant industrial interest in pressure sensors in high temperature extreme environments. MEMS pressure sensors are typically based on measurement of the displacement of a cantilever, as used for accelerometers, but with atmospheric pressure bearing on one face of the cantilever. The harsh Venus environment clearly will give rise to packaging challenges similar to those discussed in Section 2.1, which should be investigated in coming years.

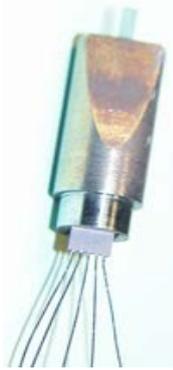

**Figure 4** An integrated 460°C weather probe which includes thermal wind sensing, as well as pressure and air temperature measurement [Hunter et al., 2009]

As to wind measurement, there are several options, all of which have significant heritage. The Russian Venera 9 and 10 probes carried spinning-cup anemometers, very much like those commonly encountered on Earth – these are best avoided on a seismometry mission because their spinning cups will cause undesirable vibrations [Avdueskii et al 1977]. The Venera 13 and 14 landers carried microphones; the noise level on the microphones was interpreted as being indicative of wind speed [Ksanfomalitii et al 1982]. Another approach is thermal anemometry, as has been used at Mars; a "Venus integrated weather sensor" measuring wind using this technique (as well as pressure and temperature) has been developed at NASA Glenn Research Centre [Hunter et al. 2009]. There has been no recent Venus-specific instrumentation work in this field in Europe, but this work could be started in time for a post-2030 mission, or a contributed instrument package from outside Europe could be solicited.

Physical accommodation of pressure sensors should be simple, as they can be mounted anywhere in the body of the lander. Accommodation of wind sensors is of course more tricky as it requires exposure to the wind, so the lander design should accommodate this requirement. Air temperature measurement is also potentially difficult; as will be shown we propose an RTG-powered craft which will dump 500 W of waste heat into the environment; an air temperature sensor will have to be as remote as possible from the RTG heat source. Data rate from a meteorological suite could be sampled at a lower rate than that of the seismometer; for datarate planning purposes we can propose that the meteorological channels are sampled at 0.1x the rate of the seismometer axes.

## 2.3  Additional payloads

The core payload is as above: a 3-axis seismometer, and minimal meteorological measurements to eliminate possible noise sources. We note that there are numerous other payloads which could be considered, once the core function of seismometry has been taken care of. Here we will list only three such possibilities: heat flux measurement, gas sensors and radio science.

**Heat flow:** The discussion in Section 1 makes clear that the rate and style by which Venus is losing its heat is a question of critical importance for understanding planetary internal structure and evolution, and so is very complementary to seismological investigations. Indeed, this has motivated the selection of the Heat flow and Physical Properties Package ($HP^3$) on Mars InSight, alongside its seismometers. Measuring heat flow on Venus should be simpler than measuring it at Mars, in that there is little or no diurnal temperature variation, so it should be possible in principle to measure the heat flow without burying the sensors; indeed, an instrument has been proposed to measure Venus' heat flow using just this principle (Pauken et al, 2016). More development and testing is needed to see whether this measurement is viable; if such testing is concluded positively we would welcome this as a US-contributed payload. It would have a low mass (<< 1 kg) and very low datarate requirement.

**Gas sensors** are an area which has seen large industrial interest, for application in internal combustion engines and turbines, as well as petroleum industry applications. An array of SiC sensors, with different active elements, can be tuned so that each has a different sensitivity to a suite of target gases; the combined response of all sensors can be read as a "fingerprint" characteristic of different gases. Once a more powerful analysis suite (employing mass spectrometers, for example, or tunable laser spectrometers) has characterised performed a detailed analysis of the near-surface atmospheric composition, a SiC gas sensor suite could track variations in the abundances of different species.

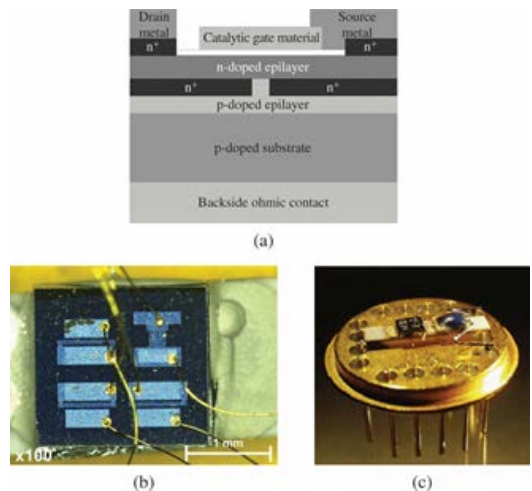

**Figure 5** High temperature gas sensors demonstrated in SiC [Bur et al. 2012]

For Venus, the application is clear: does the atmospheric composition vary at the landing site, during the months of the VL2SP's surface mission? Variations in abundances of $SO_2$ and $H_2O$ could be indicative of volcanic emission nearby; variations in carbonyl sulphide (OCS) or carbon monoxide (CO) may be indicative of surface-atmosphere reactions above specific nearby geological features.

However, there's a problem: wind speeds in the lowest 10 km of the atmosphere are very low, typically < 1 m/s, so gas abundances are expected to vary very slowly, if at all. So there is a real risk that no variations would be found. However, industrial development in this field in proceeding rapidly, and the resource requirements are low (indeed, single-chip SiC sensors are being developed for high temperature applications), so SiC gas sensors may be considered for inclusion in VL2SP pending their future development.

Finally, we should mention the value of **Radio Science** measurements, i.e. the detailed tracking of the frequency of the radio signal from VL2SP. If the VL2SP telecommunications package has a very stable frequency reference, then the Doppler shift of its signal can be tracked to precisely measure its line-of-sight velocity from either Earth or from a relay orbiter. Venus is expected to experience far greater Length-of-Day variations, and even spin axis variations, than Earth or Mars, due to its massive atmosphere. Weather variations may cause its Length of Day to vary by seconds or even minutes (Cottereau et al. 2011). Tracking these variations can yield information about the angular momentum and thus the internal structure of Venus, and is thus very complementary to the science objectives goals of VL2SP.

Conducting radio science with VL2SP would require either a 2-way telecommunications link, or an ultrastable onboard oscillator. Quartz oscillator packaging and operation at Venus temperatures was demonstrated in a NASA-funded project in 2006 [Sariri et al. 2006], but the stability of this oscillator was quoted as being in the hundreds of ppm, which is not sufficient to allow radio science investigations. Further research in this field would be welcome.

In summary, the core science payload, consisting of a 3-axis seismometer and supporting meteorological data, is expected to be less than 1 kg in weight, and 1 litre in volume, thanks to extensive use of dedicated integrated circuit solutions. The minimum required data rate is 64 bits/sec for a 1Hz sampling mission; a higher data rate of 640 Hz would allow 10 Hz seismometry and more ancillary data.

What can be achieved will clearly depend on what is achievable in high temperature electronics; we will therefore now review the current status of key high-temperature electronics elements.

# 3 High temperature electronics – current status of development

## 3.1 Background

This proposal is based on the idea that a Venus Lander could be built with all semiconductor electronics made in Silicon Carbide (SiC). Electronics circuits and discrete devices in this semiconducting material have been shown to work at 600°C by several actors [Zetterling, 2015]. Radiation hardness has also been proven for devices and circuits. The question at hand is whether the technology is mature enough for mission planning in five years (we believe so), if all building blocks can be demonstrated (we will list status) and if sufficient integration level can be achieved for all the building blocks needed (we will show some different scenarios). Finally we will list the main European and international actors in the field.

## 3.2 Maturity of SiC

Three things need to be considered regarding maturity of SiC for building an all-SiC Venus lander: availability and maturity of SiC wafer material; state of the art in SiC device and circuit processing; and commercial availability of SiC electronics. An important choice is also whether bipolar technology or field effect technology (MOSFET or JFET) should be used (we prefer the first mentioned). For background and references, see Kimoto et al. 2014 and Zetterling, 2015.

First of all it can be easily established that SiC integrated circuits are not commercially available. However, SiC high voltage switching devices have been commercially available since ten years, and can be procured from Farnell/Element 14 and other resellers. The reason is that SiC initially demonstrated its capability as a material for low loss high voltage switching devices (at normal operating temperatures) and its radiation hardness or high temperature capabilities have not yet been taken advantage of. The high voltage devices available are rated at 600 – 1700 V, and importantly for currents in the range 5 – 100 A. To make a 100 A switching device in SiC, the complexity is actually larger than the three external terminals (source, drain, gate) imply, since the current needs to be conducted by an area approaching one $cm^2$ (100 $A/cm^2$ is a typical current density in silicon power devices). The device chips used by SiC suppliers typically have higher current density, but still use around 5 x 5 mm large chips. This shows that processing yield is good enough for this size. The experience from academia (KTH) is that 7 x 7 mm chips can be made without any yield issues. The complexity inside the package of a three terminal switch is also more like 1000 transistors connected in parallel. The metallization design is much like an integrated circuit with several layers of interconnects. The process technology used is the same for discrete devices and integrated circuits. Academic demonstrators are in the range of 10 – 100 devices per circuit, but in a multi project wafer run the total number of transistors are more than 1000 in a 7 x 7 mm chip. Wafer sizes commercially available today are 100 mm and 150 mm, clearly suitable for academic and commercial use. The material quality has been good enough (without pin-holes or life-time limiting defects) for more than ten years, as is evidenced by the availability of commercial SiC high voltage devices; material quality immediately shows up in the voltage rating possible.

Integrated circuits in SiC with low level of integration (10 – 100 transistors) have been demonstrated with several technologies (type of transistors): NMOS and CMOS using

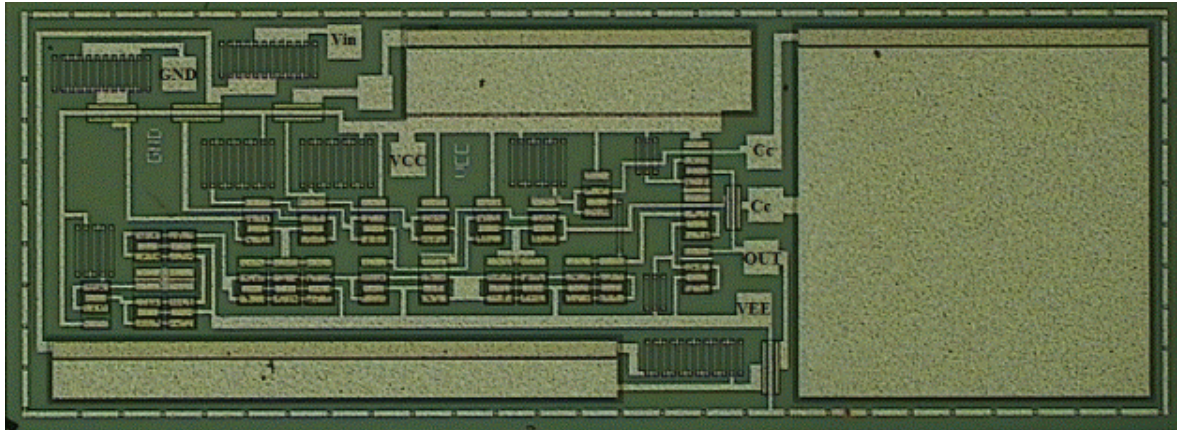

**Figure 6** Analog integrated circuit demonstrated at 500°C [Hedayati et al. 2014]

MOSFETs, MESFETs, JFETs, and bipolar transistors [Zetterling, 2015]. Of these, the reliability is the biggest challenge for MOSFETs (gate oxide integrity) and MESFETs (Schottky gate integrity), whereas neither JFETs nor bipolar transistors have specific reliability issues. All technologies require metallization that can work at extended temperature for a long time, and NASA has shown 1000 of hours in a JFET technology [Spry et al., 2016]. This metallization could be applied to any of the other, including bipolar transistors. The best reliability for radiation is expected for bipolar transistors, since they have no gate oxide and are not field effect controlled. Preliminary results show that the KTH bipolar technology can handle fluences of up to $10^{13}$ cm$^{-2}$ of 3 MeV protons and doses up to 332 Mrad of gamma radiation [Suvanam et al., 2014]. In the following we will concentrate on the KTH bipolar technology, but also show state of the art for other technologies where applicable.

### 3.3 Current status

A general electronic measurement system needs several electronic building blocks. Some solutions for a Venus lander are proposed under scenarios below, and we can list the required functions being sensors, amplifiers, analog-to-digital conversion, computing and memory, radio transmission and reception, power supply.

**Sensors**

The seismometer and other sensors have been described in section 2. The seismometer element itself is made of silicon, but its electronics will be utilise SiC components.

**Amplifiers**

Several general purpose operational amplifiers have been demonstrated up to 500°C [Hedayati et al., 2014, Tian et al., 2016]. However, to optimize the signal to noise ratio for a specific sensor, the amplifier should be custom designed to match impedances. Seismometers can be connected either open-loop or closed loop. The latter allows more sensitive measurements, but requires co-design of MEMS seismometer and amplifier. Work in this direction is being pursued at KTH within the Working on Venus project.

**Analog-to-digital conversion (ADC)**

So far a complete analog-to-digital converter has not been demonstrated in SiC, but it is being pursued at KTH in the Working on Venus project. Both flash converters and successive approximation (SAR) type are being designed. However, most of the building blocks required have been demonstrated previously at KTH in the HOTSiC project: bandgap voltage references [Hedayati et al., 2016], comparators, digital-to-analog converter (DAC) [Hedayati et al., 2016] and amplifiers. The work in the coming years is aimed at 8 bit resolution. The integration level needed for flash ADC is on the limit as the number of comparators increases as 2^n where n is the number of bits, whereas SAR type ADC is possibly limited by thermal noise in the comparator and DAC.

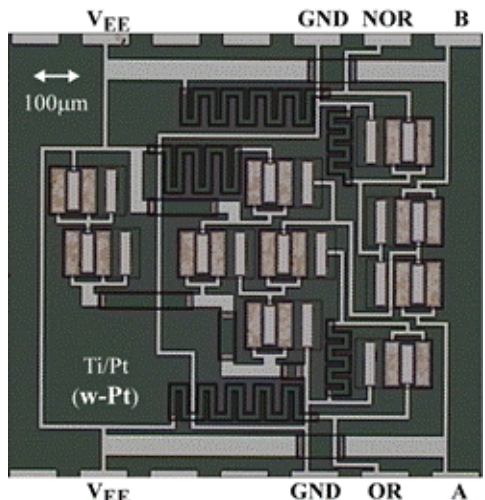

**Figure 7 SiC** Digital logic integrated circuit operated at 500°C [Lanni et al. 2015]

**Computing and memory**

A general purpose programmable microcontroller is fairly straight forward to design, and is completely digital. Digital electronics has been demonstrated up to 600°C by several groups [Patil et al., 2009, Lanni et al., 2015]. Designs for a 4 bit demonstrator with around 5000 transistors is ongoing at KTH. With a microcontroller unit (MCU) and memory data can be collected from the sensors, and filtered for events to be transmitted. Basic error correcting codes can be added (parity). Some programmability could be added in terms of changing sampling rates or sampling at specific times, with commands received on the radio. The integration level is high enough for a Venus lander microcontroller, the bottle neck is the high temperature memory. Static random access memory (SRAM) is straightforward to design, but requires 6 transistors per bit. At data rates of 1000 bits/second several kilobytes are needed, and this will have to be split on several chips. Non-volatile memory is another alternative, some initial results on ferroelectric materials that maintain their ferroelectricity at 460°C have been achieved, and the density is higher for these (1 transistor per bit). However, total memory sizes is expected to be less than 1 MB.

**Radio transmission and reception**

The amplifier for the radio transmitter needs to have high enough operating frequency for the 400 MHz band. This requires transistors with high enough gain at these frequencies, and the temperature of 460 °C. The cut-off frequency is also dependent on parasitic capacitances and inductances resulting from the layout of the transistor and its parallel connection. Fortunately the parasitic properties are not temperature dependant. SiC MESFETs and SiC SIT (static induction transistors) have been operated in L and S bands (1 – 4 GHz) but their temperature performance is not well known [Kimoto et al., 2014]. Bipolar transistors have not yet been operated at these frequencies, but their temperature dependence is known, the gain drops about 50 % from room temperature at 300°C and then levels off or even improves. The simultaneous measurement of high frequency and high temperature performance is challenging, since most high frequency probes can only handle 250 °C. Work is being pursued at KTH in the Working on Venus project to demonstrate basic radio circuits at high temperature. A backup solution is to use GaN MMICs for the radio transceiver.

**Power supply**

The power supply regulates the voltage from the radioisotope thermoelectric generator (RTG). This is typically built from discrete SiC transistors. Several SiC transistor types are commercially available, although their commercial packaging is not operative at 460 °C. Discrete transistors have been tested by KTH and typically the bipolar transistor does not degrade above 300 °C. A larger challenge is high temperature decoupling capacitors with enough capacity, but if the RTG delivers a stable output voltage this could be handled. Linear voltage regulators have been demonstrated in SiC operating at 500°C [Valle-Mayorga et al., 2014, Kargarrazi et al., 2015].

**Packaging and module assembly**

Commercially used plastic packaging and circuit boards can't be used at 460 °C. Some ceramic packages for integrated circuits might be usable at these temperatures since the ceramic firing and metal lead assembly is performed at 700°C or higher. However, since normal circuit boards can't be used, hybrid module technology is a better bet. This uses alumina ($Al_2O_3$) substrates with Ag/Pd screen-printed for connections, and wire bonding of chips directly on the substrate without packaging. This type of thick film technology is established in power electronic modules, but has not been tested yet at elevated temperatures. Preliminary work is being pursued at KTH in the Working on Venus project.

## 3.4 Expected scenarios

Based on present availability of demonstrated electronic building blocks we list three scenarios that would all enable seismometric measurements from the surface of Venus over extended time to be transmitted, with or without relaying, back to Earth. They range from the extremely simple to the more complex, and offer more or less features [based on Zetterling, 2012; similar also to options listed in KISS report, 2015].

**Scenario 1**: Direct transmission of analog signals from sensor without storage

The seismometric sensors are directly coupled to the radio transmitter and the information can modulate the amplitude of the local oscillator. In this case one transmitter is needed for each sensor, so three for a 3-axis seismometer. The advantage of this system is its simplicity, but the disadvantage is the requirement of three transmitters and three antennas. The signal to noise ratio is also poor since the sensor signal is not digitized.

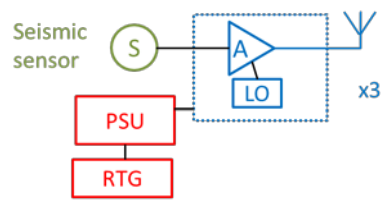

**Figure 8 -** Block diagram for VL2SP scenario 1

Key: S = Sensor, A = Amplifier, LO = Local Oscillator, PSU = Power supply unit, RTG = radioisotope thermoelectric generator (Antenna indicated with antenna symbol).

**Scenario 2**: Direct transmission of digital signal from several sensors without storage

In this scenario the sensor signals are amplified and digitized with analog-to-digital converters (one per sensor). A simple sequencer that transmits the digital information serially, one sensor at a time (and adds some parity and identifier bits) is used to collect data, but there is no storage. The advantages over scenario 1 is that the signal to noise ratio can be much better, depending on the resolution of the ADCs, and that one radio transmitter and antenna can be used for all sensors.

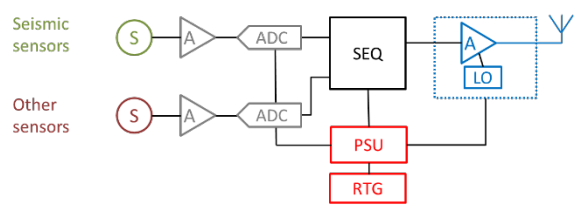

**Figure 9 -** Block diagram for VL2SP scenario 2

Key: S = Sensor, A = Amplifier, ADC = Analog-to-digital converter, SEQ = Sequencer, LO = Local Oscillator, PSU = Power supply unit, RTG = radioisotope thermoelectric generator, (Antenna indicated with antenna symbol).

**Scenario 3**: Delayed transmission of digital signal from several sensors with storage

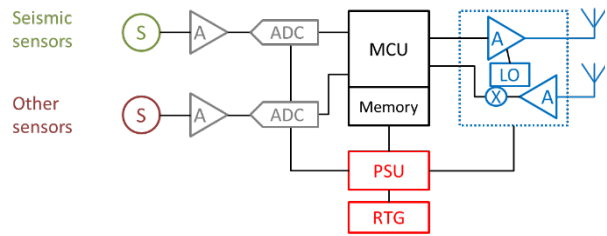

**Figure 10 -** Block diagram for VL2SP scenario 3

Key: S = Sensor, A = Amplifier, ADC = Analog-to-digital converter, MCU = micro control unit with memory, LO = Local Oscillator, PSU = Power supply unit, RTG = radioisotope thermoelectric generator, (Antenna indicated with antenna symbol).

In the third scenario more functionality is added by putting a microcontroller in the system. Not only does it sequence and error code signals, it can also use its memory to resend data, or filter data (send only interesting events). Compressing can be performed in terms of removing duplicate data, but the capacity is not that of a DSP. Depending on the size of memory it could wait before sending data. Programmability is also offered with a receive channel, that could turn on or off sensors or change sampling rates temporarily.

A note on the local oscillator in all three scenarios: The stability of oscillators in Venus conditions is not yet well characterized. Some prototypes operating in Venus conditions have demonstrated stabilities of a few hundred ppm (Sariri et al 2006) but for now we assume that the oscillator is not stable or well determined. The Earth or relay receiver will have to find and track the transmitter frequency. However, for scenario 3, since the same local oscillator is used for the Venus lander's transmitter and receiver, a protocol for transmission can be used where the signals to the Venus lander has a fixed offset from the frequency last transmitted on.

## 3.5 Capacity in Europe and beyond

Here we outline the main research institutions around the world who would be implicated in a concerted development effort towards VL2SP electronics capability.

**KTH Royal Institute of Technology, Sweden**

University research on process technology for SiC power devices since 1992, added high temperature SiC integrated circuits 2009. In-house cleanroom and bipolar technology on 100 mm wafers. High temperature testing on wafer up to 600 °C. The HOTSiC project 2011-2016 investigated basic analog and digital integrated circuits up to 500°C (www.hotsic.se). The Working on Venus project 2014-2018 aims to demonstrate all electronics needed for a Venus lander, including MCU, memory and radio (www.workingonvenus.se). The project includes ESA Astronaut Christer Fuglesang, who is adjunct professor at KTH.

**Raytheon, Glenrothes, UK**

Commercial, offers CMOS process for digital and analog circuits, demonstrated up to 400°C operation [Clark etal., 2011], some long term testing [Young et al., 2013].

**Other European active institutes:**

Newcastle University, uses Raytheon as a foundry for high temperature integrated circuits. University of Warwick, mainly high temperature power devices. Ampere Laboratory, INSA Lyon, some work on MESFET integrated circuits together with IMB-CNM Barcelona [Alexandru et al., 2015].

**CREE Research (now Wolfspeed), USA**

Commercial, pioneer in SiC material sales 1989, commercial power diodes and MOSFETs, offers NMOS process demonstrated up to 500 °C.

**NASA Glenn Research Center, USA**

Pioneer in high temperature integrated circuits, in-house JFET [Neudeck et al., 2009] and MESFET technology [Chen et al., 2006], analog and digital circuits demonstrated up to 500 °C, some long-term testing [Spry et al., 2016]. High temperature metallization and 800°C pressure sensors [Okojie et al., 2015]. Also has a Venus Environment Test chamber for environmental exposure testing.

**University of Arkansas, USA**

Integrated circuit design for high temperature [Cressler et al., 2013], uses Raytheon and CREE as foundry. CMOS from Raytheon tested up to 540°C [Rahman et al., 2016].

### 3.6 Summary and outlook

All electronic subsystems can be made in SiC, and they are scheduled to be demonstrated separately in the KTH project Working on Venus by 2018 (TRL 4). Five years from now (2021), with extended funding, improved subsystems working together and tested in a relevant atmosphere (a "Venus test chamber") could be achieved. The integration level can be increased from 1000 to 5000 transistors now to 10 000 to 50 000 transistors in five years.

## 4 Scenarios for the VL2SP element

The review of high temperature electronic components above has shown that, by early 2020s, we can expect to have op amps, comparators and ADCs, and simple 8-bit CPUs. However, memory is one critical element where high-temperature electronics are lagging; the text above shows that achievable system memory may be on the order of $10^4 - 10^5$ bits. This is insufficient for extensive storing and forwarding data, so it suggests a strategy of "live" broadcasting of all science data in near-real time. This corresponds to scenarios 2 or 3 as described in Section 3.4 above.

In this section we will first discuss power sources and then outline a possible system design for the surface element.

### 4.1 Power

It is a cruel irony that, despite its close proximity to the sun, the surface of Venus is so dimly lit. Although solar irradiance incident on the cloud-tops of Venus is 2.5 kW m$^{-2}$, most of this is scattered away by cloud droplets or Rayleigh scattering and only a few W m-2 reaches the surface. Although it may be possible to develop solar panels which operate at Venus temperatures, the low level of sunlight means that impractically large arrays would be required.

Primary (non-rechargeable) batteries could be considered for missions of short duration, and indeed could be used for a pathfinder mission whose aim is to establish levels of seismic activity on Venus as a precursor to a more ambitious seismometer mission. A recent study [Salazar et al., 2014] concluded that primary batteries were viable for missions of less than a few months in duration.

For missions lasting longer than a few months, then, we must consider radioisotope powered sources of electricity. The simplest solution is to use a radioisotope thermoelectric generator (RTG), which has no moving parts and is therefore ideal for a seismometry mission.

Salazar et al (2014) found that an RTG of 24 kg could deliver electric power of 26 W (as well as roughly 500 W of thermal power). This seems consistent with the power levels needed to achieve transmission rates of a few hundred bits per second from the surface of Venus to an orbiter, so is assumed as the baseline in our study. We (the author team) are not aware of specific RTG developments in Europe which could be used instead, so are using the results of this U.S. study as a sizing case.

Alternatively, higher powers (and higher costs!) can be achieved using Stirling Cycle converters. The reciprocating piston in these devices may introduce spurious signals into the seismometer data; mechanical and/or electrical filtering would need to be used in order to minimise the interference with seismic data (see e.g. Lorenz et al, PSS 2012, for further discussion).

## 4.2 Mechanical design of VL2SP element

We have not conducted any design work or conducted any trade-off studies so at this point we can offer only some constraints.

First of all, the VL2SP has to house the seismometer itself, as well as the RTG power source, telecomms system and associated electronics.

The RTG generates 500 W of heat which must be dissipated; this should be lost to the atmosphere through natural convection using fins.

The seismometer should be well coupled to the ground, such that it is does not wobble or sway when the wind blows. One way to achieve this is to mount the VL2SP on a rigid tripod; however this carries the risk that the seismometer may be balanced atop a rock if the seismometer lands on an uneven surface. Another approach may be to have a crushable conformal structure, such as a honeycomb, which allows the bottom surface of the VL2SP to reach good contact with the ground.

The lander should present the smallest possible cross-section to wind, to minimise wind stress. Wind speed will increase with increasing distance from the surface, so keeping the lander very low in height is one way to address this; deployment of a "wind shield" would be another. Burying the seismometer would of course be another way to guarantee good coupling to the surface, and this could be achieved with a penetrator – but the thick atmosphere of Venus and presumed

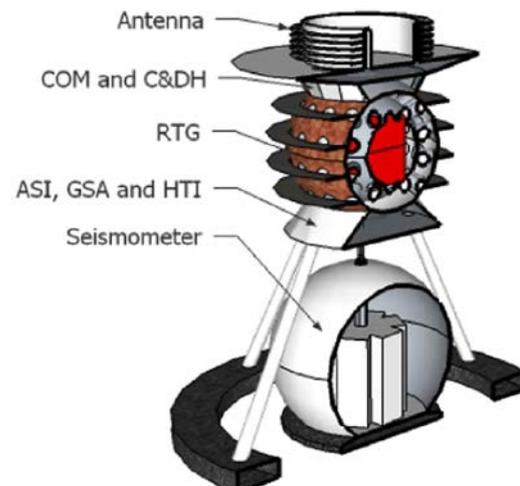

**Figure 11** HADES lander concept, seismometer is under a wind shield [Boll et al. 2015]

hard basaltic surface of makes it difficult to propose a penetrator mission.

This leads us to propose two scenarios. One scenario is as proposed by Boll et al (2014), as shown in Fig. 11: a Venera-style lander descends and lands on a landing ring; a central spherical enclosure holds RTG and electronics; the seismometer and wind shield are lowered to the surface to be mechanically independent of the lander. However, this design may be difficult to realise in practice: for example, lock and deployment mechanism for the seismometer are non-trivial; also, the fairly tall lander structure with its shock absorber legs would sway in the wind and may transmit some of these vibrations to the surface below where the seismometer sits.

A simpler solution may be as shown in Fig. 12: If the seismometer and RTG are mounted in a monolithic, disc-like structure, then it is ensured that it will land flat on one of its two faces. A crushable conformal structure (e.g. aluminium honeycomb) on both faces would seek to ensure good contact to the ground; the low height of the structure would ensure that it presents a low profile to the wind. A potential problem is compatibility with the antenna required for a communications system; potentially, a patch antenna could be mounted on both of the faces, and the selection between antennae made by examining an accelerometer to determine the direction of the local gravitational vector. The choice of mechanical design approach will clearly depend on the mission scenario, i.e. whether the VL2SP is delivered to Venus as a passenger of a larger mission: this is discussed in the following section.

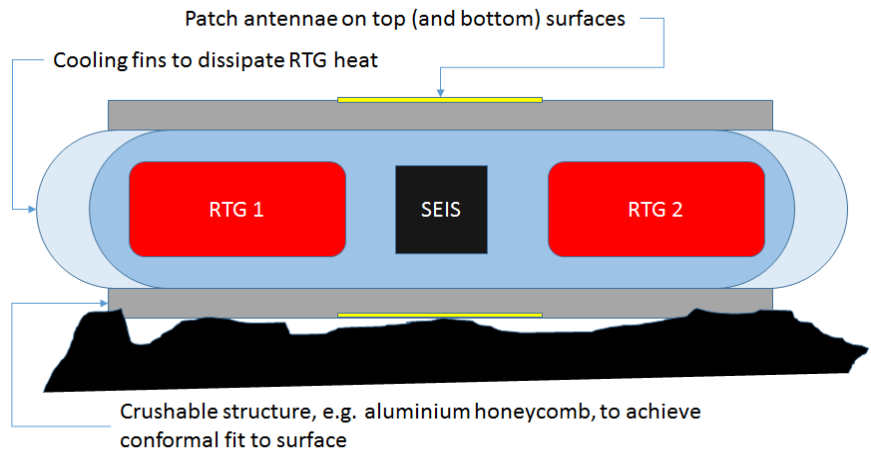

**Figure 12 -** Sketch of monolithic disc-like VL2SP design, low to the ground.

## 5 International context & mission scenarios

In the previous section, we have proposed an RTG-powered VL2SP which would broadcast its science data in near-real time. Our preliminary calculations indicate that a data relay orbiter would be needed in order to achieve the required data rates of several hundred bits per second. We must therefore either consider what orbiters might already be in place in a post-2030 timeframe, or consider a mission which delivers both VL2SP and a data relay orbiter. If VL2SP has a dedicated relay orbiter, one should consider whether this orbiter can carry a camera optimised for imaging airglow oscillations which may arise from seismic activity [Lognonné et al, 2016].

Secondly, VL2SP must be delivered safely to the surface of Venus. Again, we can either consider VL2SP as a stand-alone mission or as a secondary element attached to a larger mission. If it is a stand-alone mission, then a Thermal Protection System has to be designed to protect the VL2SP during the entry phase; the aerodynamics of the VL2SP during the descent phase should be studied to see if further deceleration devices such as parachutes are needed.

Alternatively, it might be convenient to develop the VL2SP as an autonomous element to be delivered to the surface of Venus by a Venus lander mission. Roskosmos plans a "Venera-D" lander, to launch post-2026, which would soft-land on Venus but would only last a few hours on the surface; it might be feasible to have VL2SP delivered to the surface of Venus by this vehicle. Venus surface missions are also in development elsewhere: the NASA New Frontiers call for example, specifically invites Venus In Situ Explorer missions. Finally, there may be many Venus flybys in coming years (for example, there are some ten flybys in the next decade from Solar Orbiter, Solar Probe +, and Bepi Colombo performing gravity assist manoeuvres), so one option to be considered would be a VL2SP in its own dedicated which could be delivered to Venus by a passing spacecraft.

To summarise: a stand-alone VL2SP mission would require not only the surface element but also a data relay orbiter but also an Entry/Descent/Landing system; however, studies of VL2SP as a secondary mission element would be beneficial.

# 6 Future development roadmap

*Wide bandgap semiconductors such as gallium nitride (GaN), silicon carbide (SiC) and diamond have emerged as some of the most promising materials for future electronic components. They offer significant advantages in terms of power capability (DC, power switching and microwave), robustness against radiation, high temperature and high frequency operation, optical properties and even low noise capability. Therefore wide bandgap components are strategically important for the development of next generation space and defence systems.*

The above text was taken from the website of an ESA conference being held as we write this document[2]. This bears witness to the widespread application of such electronics systems across the space sector; when additional applications in geophysical, automotive and aviation applications are taken into account, it is clear that there are many potential applications of developments in this field. The present VL2SP application exploits developments in this field, but also acts as an added motivation to drive new development. For telecommunications applications, for example, there has been a focus on component-level amplifiers and DC-DC converters for high power, high temperature or high radiation designs. The VL2SP proposal seeks to generalise these developments to increase the complexity of components which can be implemented in SiC, which will have benefits for the telecomms sector.

For a VL2SP mission to be feasible in a 2030-2040 timeframe, several specific development programmes would be recommended:

- Continued development of the electronics components described in Section 4 of this document: ADCs, digital logic, volatile and non-volatile memory, and telecommunications circuits.
- Packaging for high-temperature electronics. Interconnect techniques and encapsulation materials.
- High temperature sensors: seismometer; meteorological sensor suite; ultra-stable oscillator.
- A small (1-20 L) Venus environment test chamber, containing $CO_2$ at 90 bar and 460 °C, to test exposure of all of the above components to Venus surface conditions. We propose a smaller, cheaper chamber than the one at NASA Glenn, to allow rapid component-level testing.

---

[2] 8th ESA Wide Bandgap Semiconductors Conference, Harwell, 12-13 Sep 2016.

Venus Long-Life Surface Platform (VL2SP)                                                                      Page 18/19startrefsend